# The occupation dependent DFT-1/2 method

Shengxin Yang,[1] Jiangzhen Shi,[1] Kan-Hao Xue,[1,*] Jun-Hui Yuan,[2] and Xiangshui Miao[1]

[1]School of Integrated Circuits, Huazhong University of Science and Technology, Wuhan 430074, China
[2]School of Physics and Mechanics, Wuhan University of Technology, Wuhan 430070, China

*Corresponding Author, Email: xkh@hust.edu.cn (K.-H. Xue)

## Abstract

There has been a high demand in rectifying the band gap under-estimation problem in density functional theory (DFT), while keeping the computational load at the same level as local density approximation. DFT-1/2 and shell DFT-1/2 are useful attempts, as they correct the spurious electron self-interaction through the application of self-energy potentials, which pull down the valence band. Nevertheless, the self-energy potential inevitably disturbs the conduction band, and these two methods fail for semiconductors whose hole and electron are entangled in the same shell-like regions. In this work, we introduce the occupation-dependent DFT-1/2 method, where conduction band states are not subject to the additional self-energy potential disturbance. This methodology works for difficult cases such as $Li_2O_2$, $Cu_2O$ and two-dimensional semiconductors. Using a shell-like region for the self-energy potential, and allowing for downscaling of the atomic self-energy potential (with an $A < 1$ factor), the occupation-dependent shell DFT+$A$-1/2 method yields more accurate conduction band and valence band edge levels for monolayer $MoS_2$, compared with the computationally demanding hybrid functional approach.

## I. INTRODUCTION

The band gap problem [1] is a well-known issue in applying density functional theory [2,3] (DFT) to the calculation of semiconductor electronic structures. In average, the band gap is subject to ~40% underestimation for semiconductors and insulators, under local density approximation [3–7] (LDA) calculations. Intrinsically, this is because the Kohn-Sham eigenvalues are physically not linked to experimental accessible quantities. Even the energy eigenvalue of the valence band maximum (VBM) is far from minus the work function in practical calculation, apparently in contradiction to a well-known theorem [8–10] of DFT. This is because the theorem stating that the highest occupied energy level is equal to minus the work function has a prerequisite condition: the exchange-correlation (XC) functional is exact. In practical LDA or generalized gradient approximation (GGA) calculations, the incomplete cancellation of the spurious electron self-interaction [11] renders too high VBM levels [12], accounting for most of the band gap under-estimation. In 2008, Ferreira, Marques and Teles [13] proposed the innovative DFT-1/2 method for solid-state calculations, which rectifies the electron self-interaction problem through introducing the so-called "self-energy potential" (SEP). The SEPs are included for those "anions" that contribute to the valence band, thus DFT-1/2 yields lower VBM levels and much improved band gaps for semiconductors and insulators [14,15]. In the mean time, its computational complexity is still kept at the LDA/GGA level [16], and self-consistent cycle must be enforced to generate the band structures [17].

Later, it was found that DFT-1/2 has some problems for certain covalent semiconductors, and the shell DFT-1/2 method was proposed in 2018 [18]. Their difference only lies in the way of trimming the SEPs. Standard DFT-1/2 uses a spherical trimming function, and the cutoff radius $r_{cut}$ is optimized to maximize the band gap, which is theoretically equivalent to a minimum ground state energy. On the other hand, shell DFT-1/2 allows for shell-like regions to filter out the valence band hole location [19]. Hence, in shell DFT-1/2 there are two cutoff radii, $r_{in}$ and $r_{out}$, which account for the inner limit and outer limit of the shell, respectively. Take GaAs as an example. In DFT-1/2, the sole anion is As, and a spherical trimming for the As SEP yields a satisfactory $\Gamma - \Gamma$ band gap of 1.56 eV. In shell DFT-1/2, analyses of the conduction band electron and valence band hole spatial distributions imply that it is shell DFT-1/4-1/4 that has to be implemented. This means that both Ga and As are subject to -1/4 e correction, but their inner cutoff radii of the SEPs are not zero. Variational calculation yields the shell region [2.1, 3.9] Bohr for Ga, and [1.4, 3.2] Bohr for As. The resulting shell DFT-1/2 band gap is $\Gamma - \Gamma$ with a value of 1.54 eV [18]. Yet, the indirect gaps at L and near X are more accurate by shell DFT-1/2. Moreover, shell DFT-1/2 predicts Ge to be an indirect $\Gamma - L$ semiconductor, while standard DFT-1/2 predicts a direct gap. The reason lies in that, the near-core region sees high probability for the conduction band electron [18]. In standard DFT-1/2, the energy level of the conduction band is unexpectedly pulled down by the Ge SEP, and the energy eigenvalue at $\Gamma$ drops faster than L [20].

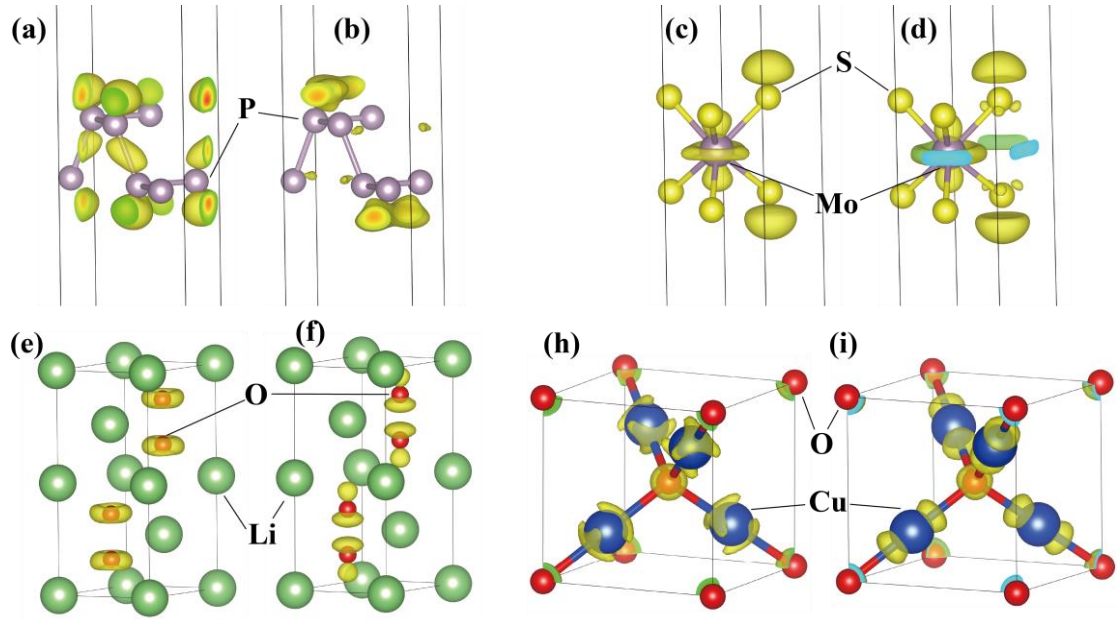

**Figure 1**. Spatial distributions of valence band hole and conduction band electron in several materials. (a) Hole of monolayer BP; (b) Electron of monolayer BP; (c) Hole of monolayer $MoS_2$; (d) Electron of monolayer $MoS_2$; (e) Hole of $Li_2O_2$; (f) Electron of $Li_2O_2$; (g) Hole of $Cu_2O$; (h) Electron of $Cu_2O$.

While shell DFT-1/2 could bring about some improvement in better filtering out the hole in real space, through implementing the shell-like trimming function, it still has intrinsic deficits. Its success relies on the fact that the hole and the electron are distributed in different shell-like regions, but there are semiconductors whose hole and electron are entangled in the same spherical shell. **Figure 1** illustrates the most probable hole and electron spatial locations in a two-dimensional (2D) semiconductor black phosphorus (BP). A shell-region cannot cover the hole while not severely disturbing the electron. Since shell DFT-1/2 attaches the SEPs in the pseudopotentials, they equally influence any state that has certain probability to emerge in the SEP regions. Consequently, the shell DFT-1/2 band gap for BP is supposed to be severely underestimated. The same problem is encountered in certain transitional metal oxides such as $Cu_2O$ [21].

In order to avoid the failure of shell DFT-1/2, one has to judge whether a state should be influenced by the SEP. Hence, the ultimate solution is to do an occupation-dependent DFT-1/2 calculation, which guarantees that the empty conduction band should not be erroneously pulled down by the SEP. Hence, it is the aim of this paper to explore such an occupation-dependent DFT-1/2 method, which may be called "occ DFT-1/2" for brevity.

## II. Principles of occupation dependent DFT-1/2

The conventional DFT-1/2 method can be regarded as introducing an extra correction term in the Kohn-Sham equation

$$[\hat{T}_s + \hat{V}_{\text{Hartree}} + \hat{V}_{\text{XC}} + \hat{V}_{\text{ps}} + \hat{V}_S]\psi_{n,k}(r) = \varepsilon_{n,k}\psi_{n,k}(r) \qquad (1)$$

where $\hat{T}_s$ is the one-electron kinetic energy operator, $\hat{V}_{\text{Hartree}}$ is the Hartree potential for the electronic system, $\hat{V}_{\text{XC}}$ is the XC potential that is available through the DFT functional in use, $\hat{V}_{\text{ps}}$ is the global pseudopotential operator that consists of contributions from all atoms, and $\hat{V}_S$ is the global DFT-1/2 correction potential that is a combination of properly trimmed SEPs. Here $\psi_{n,k}$ is the Kohn-Sham orbital wavefunction; $n$ and $k$ stand for the band and $k$ point indices for the energy eigenstate, respectively. The trimmed $\hat{V}_S$ is expressed as

$$\hat{V}_S(r) = \sum_\kappa \sum_\omega \sum_T V_S^\kappa(r - \tau_{\kappa,\omega} - T) \qquad (2)$$

where $V_S^\kappa(r) = V_S^\kappa(r) = \Theta(r)V_S^{\kappa,0}(r)$ is the trimmed SEP for element $\kappa$, and $V_S^{\kappa,0}(r)$ is the untrimmed SEP directly obtained from atomic calculations. The dummy variable $\omega$ represents all atoms with element $\kappa$ in a primitive cell, and $T$ is any translation vector for the Bravais lattice. Hence, $\hat{V}_S(r)$ involves sum over all elements $\kappa$, and all atoms with element $\kappa$ in the lattice. The trimming function $\Theta(r)$ is defined as, in conventional DFT-1/2 [13],

$$\Theta(r) = \begin{cases} [1 - (r/r_{\text{cut}})^p]^3, & r \leq r_{\text{cut}} \\ 0, & r > r_{\text{cut}} \end{cases} \qquad (3)$$

where $r_{\text{cut}}$ is the cutoff radius for the SEP. In shell DFT-1/2 [18], one has

$$\Theta(r) = \begin{cases} 0, & r < r_{\text{in}} \\ \left\{1 - \left[\frac{2(r - r_{\text{in}})}{r_{\text{out}} - r_{\text{in}}} - 1\right]^p\right\}^3, & r_{\text{in}} \leq r \leq r_{\text{out}} \\ 0, & r > r_{\text{out}} \end{cases} \qquad (4)$$

where $r_{\text{in}}$ and $r_{\text{out}}$ are the inner and outer cutoff radii, respectively. In practice, the trimmed SEPs are typically included in the local part of the atomic pseudopotentials. In other words, $\hat{V}_{\text{ps}}$ and $\hat{V}_S$ are usually treated together.

To realize occ DFT-1/2, our style of implementation is similar to DFT+$U$ [22], though the two methods are different in their physical meanings. Take the simplest Dudarev [23] rotationally invariant DFT+$U$ method as an example. Its potential operator can be written as

$$\hat{V}_U = \sum_{\tau,\sigma} \sum_{m,m'} |\Phi_m^{\tau\sigma}\rangle \left[U_{\text{eff}}\left(\frac{1}{2}\delta_{m,m'} - \rho_{m,m'}^{\tau\sigma}\right)\right] \langle \Phi_{m'}^{\tau\sigma}| \qquad (5)$$

where, $\Phi_m^{\tau\sigma}$ represents the atomic orbital, $\tau$ denotes the atomic index, $\sigma$ indicates the spin, $m$ is the magnetic quantum number, and $U_{\text{eff}}$ is the effective Hubbard $U$ [24] parameter. And $\delta_{m,m'}$ refers to the Kronecker delta symbol, while $\boldsymbol{\rho}^{\tau\sigma}$ is the occupation matrix for atomic orbitals (specifically $d$ or $f$ orbitals) at atomic position $\tau$. Its matrix elements are defined as

$$\rho_{m,m'}^{\tau\sigma} = \sum_{k,n} f_{n,k} \langle \psi_{n,k} | \Phi_{m'}^{\tau\sigma} \rangle \langle \Phi_m^{\tau\sigma} | \psi_{n,k} \rangle \tag{6}$$

Specifically, $f_{n,k}$ denotes the occupancy of the $n^{\text{th}}$ band at a given $k$ point. The potential operator within the DFT+$U$ formalism can be summarized as a two-step process: a first projection onto the atomic orbitals, followed by an on-site energy correction.

In occ DFT-1/2, we also use projectors like $|\cdot\rangle\langle\cdot|$ to determine whether the self-energy correction should be carried out for a given state. The full SEP operator is written as

$$\hat{V}_S^{\text{occ}} = \sum_{n \leq M} |\psi_{n,k}\rangle V_S(\boldsymbol{r}) \langle \psi_{n,k}| \tag{7}$$

where $M$ represents the highest occupied eigenstate. The above operator is better understood in terms of its matrix element

$$\langle \psi_{n,k} | \hat{V}_S^{\text{occ}} | \psi_{n,k} \rangle = \begin{cases} \langle \psi_{n,k} | V_S(\boldsymbol{r}) | \psi_{n,k} \rangle, & n \leq M \\ 0, & n > M \end{cases} \tag{8}$$

There is still an intricate issue here. In standard DFT code, the ground state is gradually reached through an iterative self-consistent cycle, and the ground state wavefunctions are not known in advance. Hence, the occupied states for $\hat{V}_S^{\text{occ}}$ refer to the last iterative step version $\left|\psi_{n,k}^{[i-1]}\right\rangle$. When the Hamiltonian is applied onto a trial wavefunction $\left|\psi_{n,k}^{\text{trial}}\right\rangle$, the $\hat{V}_S^{\text{occ}}$ part behaves as

$$\hat{V}_S^{\text{occ}} \left|\psi_{n,k}^{\text{trial}}\right\rangle = \sum_{n \leq M} \left|\psi_{n,k}^{[i-1]}\right\rangle V_S(\boldsymbol{r}) \left\langle \psi_{n,k}^{[i-1]} \middle| \psi_{n,k}^{\text{trial}} \right\rangle \tag{9}$$

Upon convergence in the self-consistent cycle, the errors of ground state electron density as well as the total energy are below an acceptable threshold, which renders $\left\| \left|\psi_{n,k}^{[i]}\right\rangle - \left|\psi_{n,k}^{[i-1]}\right\rangle \right\|$ also below a critical value. Hence, $\left\langle \psi_{n,k}^{[i-1]} \middle| \psi_{n,k}^{[i]} \right\rangle \approx 1$, and the occ DFT-1/2 run is regarded as reaching convergence. In this work, we have modified the ABACUS code [25,26] to enable occ DFT-1/2 run, and the Hamiltonian is solved within the plane wave-pseudopotential framework [27].

# III. Methodology to obtain the SEP cutoff radii

The above procedure is reasonable provided that one knows what cutoff radii are to be used for the SEPs. Although occ DFT-1/2 sounds reasonable and more versatile, there is a cost in forcing the occupation dependency. Traditional DFT-1/2 and shell DFT-1/2 enjoy a variational procedure to fix the cutoff radii, through maximizing the band gap. This is because outside the region that should be covered by the SEP, the conduction band electron has higher probability to emerge than the hole. Consequently, the band gap shrinks when the outer cutoff radius goes beyond the proper value (or the inner cutoff radius becomes less than the proper value, for shell DFT-1/2 only). In occ DFT-1/2, the conduction band is guaranteed not to be pulled down by the SEP. Take BP as an example. When scanning the cutoff radius in occ DFT-1/2, it is discovered that the band gap keeps on increasing upon enlarging $r_{cut}$, as shown in **Figure 2(a)**. The absence of a maximum band gap proposes a question as how to fix the cutoff radius for the SEP.

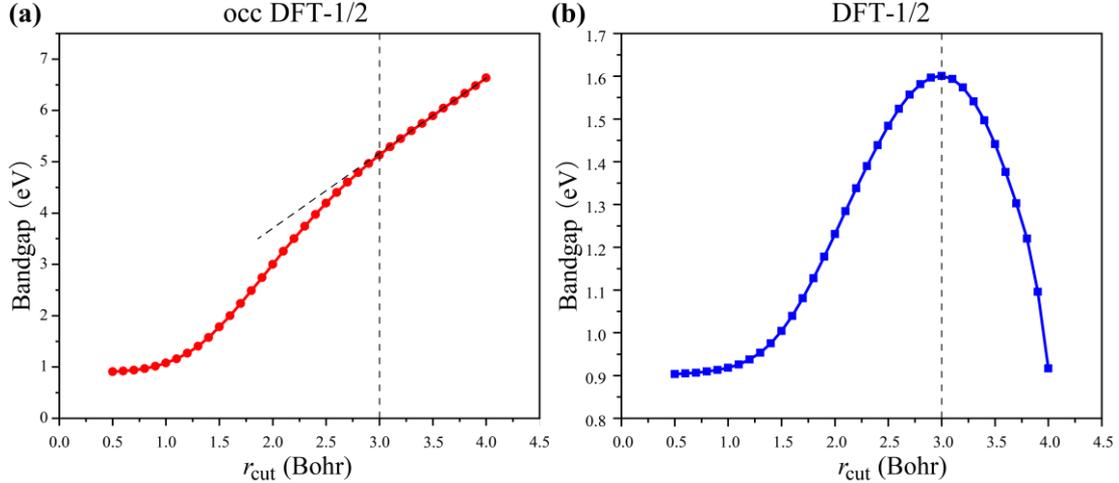

**Figure 2**. Calculated band gap of monolayer BP with respect to $r_{cut}$: (a) occ DFT-1/2; (b) standard DFT-1/2.

The reason for an ever-increasing band gap in occ DFT-1/2 is analyzed as follows. Traditional DFT-1/2 pulls down the valence band as well as the conduction band, and beyond the appropriate outer radius, it pulls down the conduction band more heavily than valence band. In occ DFT-1/2, the conduction band is not pulled down, and the SEPs will eventually cover each other since the SEP of one atom can extend to the region of another atom. The sole functionality of the SEPs is to pull down the valence band, thus the band gap does not exhibit a maximum. Yet, one may inspect the slope of the band gap variation with respect to $r_{\text{cut}}$. In **Figure 2**, it turns out that the slope becomes stable for $r_{\text{cut}} > 3$ Bohr in monolayer BP. Since the SEP is an attractive potential, it tends to withdraw surrounding electrons into the region covered by it. For this reason, the valence band is lowered when the SEP region expands due to a volume effect, even if the newly covered region has no hole distribution initially. On the other hand, we used standard DFT-1/2 to calculate monolayer BP, where the optimal $r_{\text{cut}}$ is also 3 Bohr, corresponding to the critical radius of occ DFT-1/2, beyond which the slope is constant.

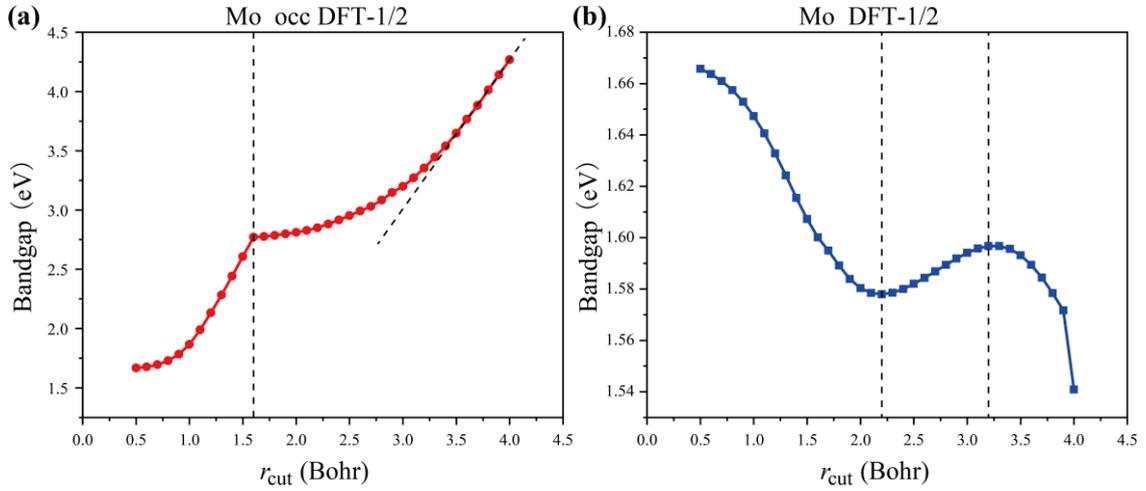

**Figure 3**. Calculated band gap for MoS$_2$ versus $r_{\text{cut}}$, with self-energy correction on Mo only: (a) occ DFT-1/2; (b) standard DFT-1/2.

Nevertheless, not every semiconductor shows such a simple trend. MoS$_2$ is probably the most important 2D semiconductor at present [28]. Its valence band hole is mainly located around Mo atoms. Carrying out occ DFT-1/2 correction for monolayer MoS$_2$, the band gap variation with respect to $r_{\text{cut}}$ is illustrated in **Figure 3(a)**. On the one hand, the band gap keeps on increasing, which is normal for occ DFT-1/2. On the other hand, the trend in the slope is quite unusual: at ~1.6 Bohr the slope suddenly drops down, but it then gradually recovers until a stable slope is reached at large $r_{\text{cut}}$. Is it reasonable to take the transition point 1.6 Bohr as the final $r_{\text{cut}}$? This answer is simply not, by referring to the DFT-1/2 result as shown in **Figure 3(b)**. As $r_{\text{cut}}$ increases from zero, the DFT-1/2 band gap of MoS$_2$ decreases monotonically until 2.2 Bohr. Yet, then the band gap increases and a local maximum is observed at 3.2 Bohr. For the $r_{\text{cut}}$ range under investigation, however, the global maximum is still $r_{\text{cut}} = 0$, *i.e.*, no correction. Explanation to this peculiar behavior is as follows. Close to the Mo core, there are conduction band electrons rather than holes that distribute. Therefore, a small $r_{\text{cut}} < 2.2$ Bohr mainly pulls down the conduction band, resulting in a smaller band gap than conventional DFT. Near 2.2 Bohr, the hole concentration becomes considerable, thus the SEP pulls down the conduction band and valence band altogether. And in case the hole concentration exceeds that of the electron, the band gap will turn to increase nevertheless. This implies that shell DFT-1/2 may be attempted. Using shell DFT-1/2 correction for Mo only, the band gap is obtained as 1.71 eV, corresponding to a SEP region of [2.3, 3.3] Bohr. The two cutoff radii are similar to the locations of the dashed lines in **Figure 3 (b)**. While the shell DFT-1/2 gap is still too small, it generates a reasonable range for the SEP cutoff radii. On account of the difficulty in obtaining the occ DFT-1/2 cutoff radii for MoS$_2$, it is quite reasonable to stick to the cutoff radius (here outer radius) obtained from shell DFT-1/2.

There are still two remaining issues for MoS$_2$. First, the hole is also localized around the S anions, and such regions are independent to the Mo SEP region. In other words, the hole is not shared as in conventional shell DFT-1/4-1/4 between Mo and S, but both Mo and S are being subject to -1/2 e correction. The optimal inner and outer cutoff radii for the S SEP are 0 and 3.2 Bohr, respectively. Second, though there is no difference whether using occ DFT-1/2 or using occ shell DFT-1/2 as far as the S correction is concerned (since the $r_{\text{in}} = 0$ for S), whether the inner cutoff should be applied remains an open question for Mo. **Figure 4(a-d)** provide the band diagrams of MoS$_2$ calculated using PBE, shell DFT-1/2, occ DFT-1/2 as well as occ shell DFT-1/2. The SEP for Mo covers a radial range of [2.3, 3.3] Bohr in occ shell DFT-1/2. It follows from **Figure 4**(a) that, PBE calculation predicts monolayer MoS$_2$ as an indirect gap semiconductor, though the maximum occupied eigenvalue at K is merely lower than that of Γ by 0.02 eV. Shell DFT-1/2 does not perform very well in terms of the band, since it is only enhanced by 0.05 eV. However, shell DFT-1/2 could correctly predict monolayer MoS$_2$ as a $K - K$ direct gap semiconductor. In terms of occ DFT-1/2, the band gap is overlarge, and the type of gap is indirect. Finally, occ shell DFT-1/2 recovers again the direct gap feature of monolayer MoS$_2$. While its band gap is still overlarge, it is much closer to the experimental value than occ DFT-1/2. Yet, the valence band of occ shell DFT-1/2 (from -1.5 eV to zero, Mo-dominated) is too far above the rest bands (those below -3 eV, S-dominated), which is a serious problem. There is no such problem in occ DFT-1/2, for the reason below. The outer cutoff radii for Mo and S are similar, and the Mo-dominated band and S-dominated band are pulled down to a similar extent according to occ DFT-1/2. Nevertheless, there is a sizable inner cutoff radius for Mo in occ shell DFT-1/2, which renders the Mo band pulled down less severely than the S-band. To sum up, both occ DFT-1/2 and occ shell DFT-1/2 are not perfect for monolayer $MoS_2$. The former leads to a mistake in the type of band gap, while the latter pulls down the S-band more heavily than the Mo-band, rendering inaccurate valence band structure.

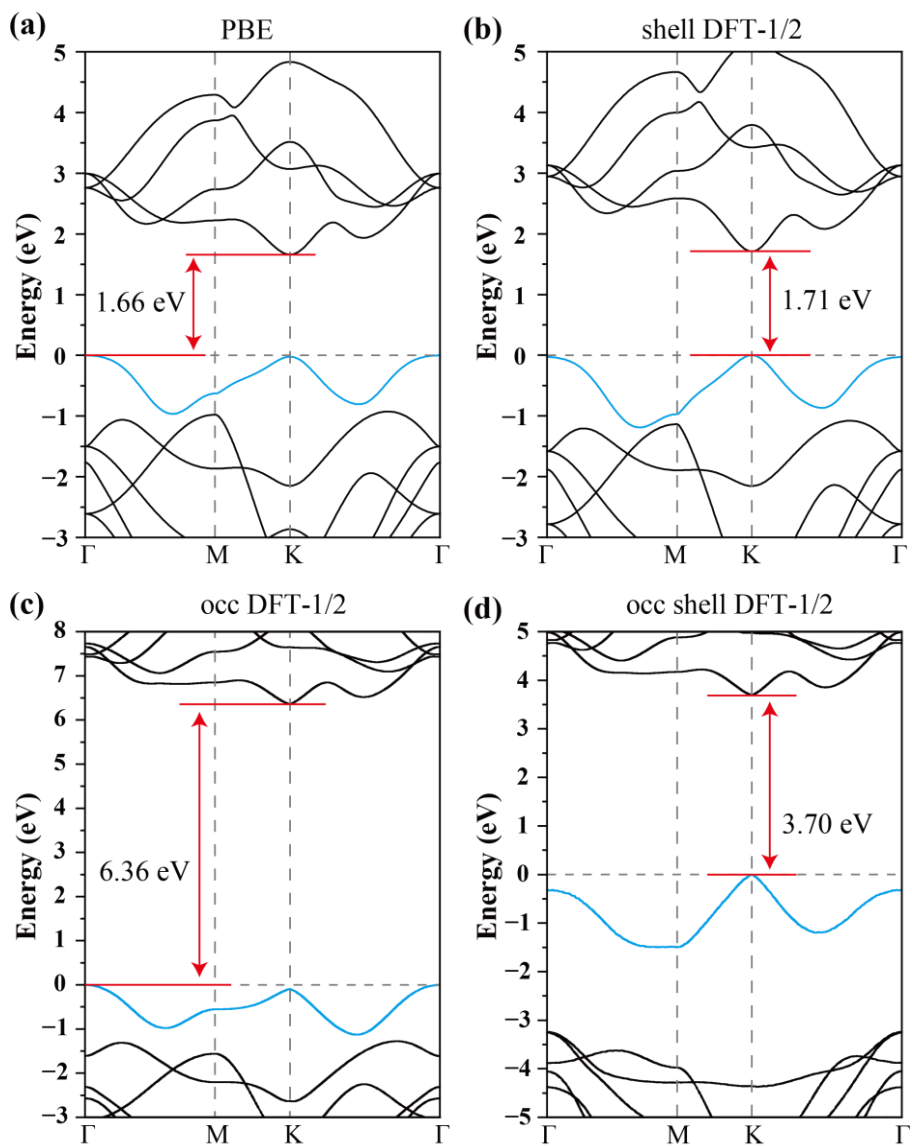

**Figure 4**. Electronic band structures of monolayer $MoS_2$ obtained using various methods: (a) PBE; (b) shell DFT-1/2; (c) occ DFT-1/2; (d) occ shell DFT-1/2.

We have also utilized shell DFT-1/2 to obtain the cutoff radii for other semiconductors under investigation. For monolayer BP, the SEP radius range is [0, 3.0] Bohr, corresponding to a shell DFT-1/2 band gap of ~1.60 eV. In $Li_2O_2$, the inner and outer SEP cutoff radii for O are 0.8 Bohr and 1.9 Bohr, respectively. The shell DFT-1/2 gap for $Li_2O_2$ is merely 2.95 eV. In $Cu_2O$ it is $Cu^+$ that is subject to self-energy correction, with a SEP radius range of [0.5, 2.5] Bohr and an shell DFT-1/2 gap value of 0.83 eV. Occ DFT-1/2 is indeed required to yield sufficiently large band gaps for these difficult cases, but the cutoff radii obtained through conventional shell DFT-1/2 are quite useful.

## IV. Amplitude modulation of the self-energy potential

The case of $MoS_2$ demonstrates that, both occ DFT-1/2 and occ shell DFT-1/2 predict overlarge band gaps. This fact may be explained through plotting the CBM and VBM levels with respect to the vacuum level as in **Figure 5**. Compared with LDA or GGA, shell DFT-1/2 involves the SEP term in the Kohn-Sham equation, which does not discriminate the conduction band from the valence band. The shell-like cutoff function is the key to filter out the valence band hole, rendering a much more severe lowering of the valence band compared with the conduction band ($\Delta E_V > \Delta E_C$). In occ shell DFT-1/2, however, whether a state is influenced by the SEP relies on its occupancy. Only the valence band is pulled down, rather than the conduction band. Since the shell DFT-1/2 band gaps are usually reasonable, the absence of conduction band downshift therefore renders extraordinarily large band gaps. If the SEPs are weakened in occ shell DFT-1/2, one may still get reasonable band gaps. Accordingly, one should ask the following question, that whose CBM and VBM are more accurate with respect to the vacuum level: (i) standard shell DFT-1/2; (ii) occ shell DFT-1/2 with diminished SEPs?

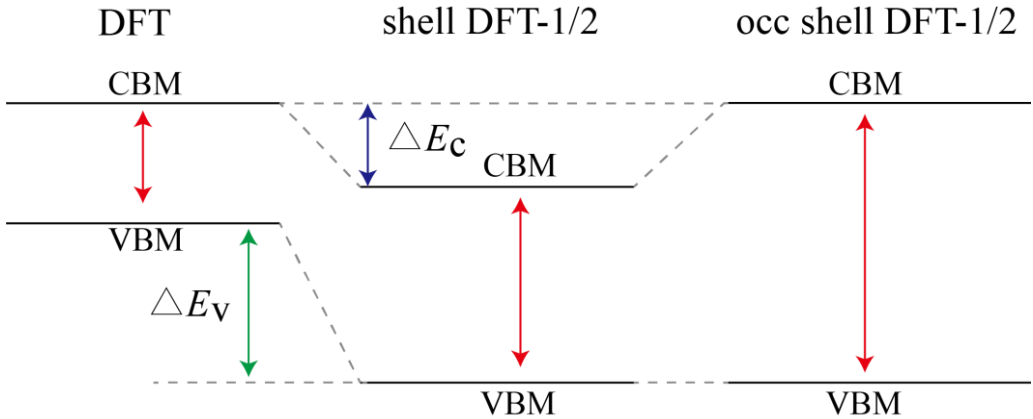

**Figure 5**. A schematic illustration of the CBM, VBM locations predicted by DFT (LDA or GGA), shell DFT-1/2 and occ shell DFT-1/2.

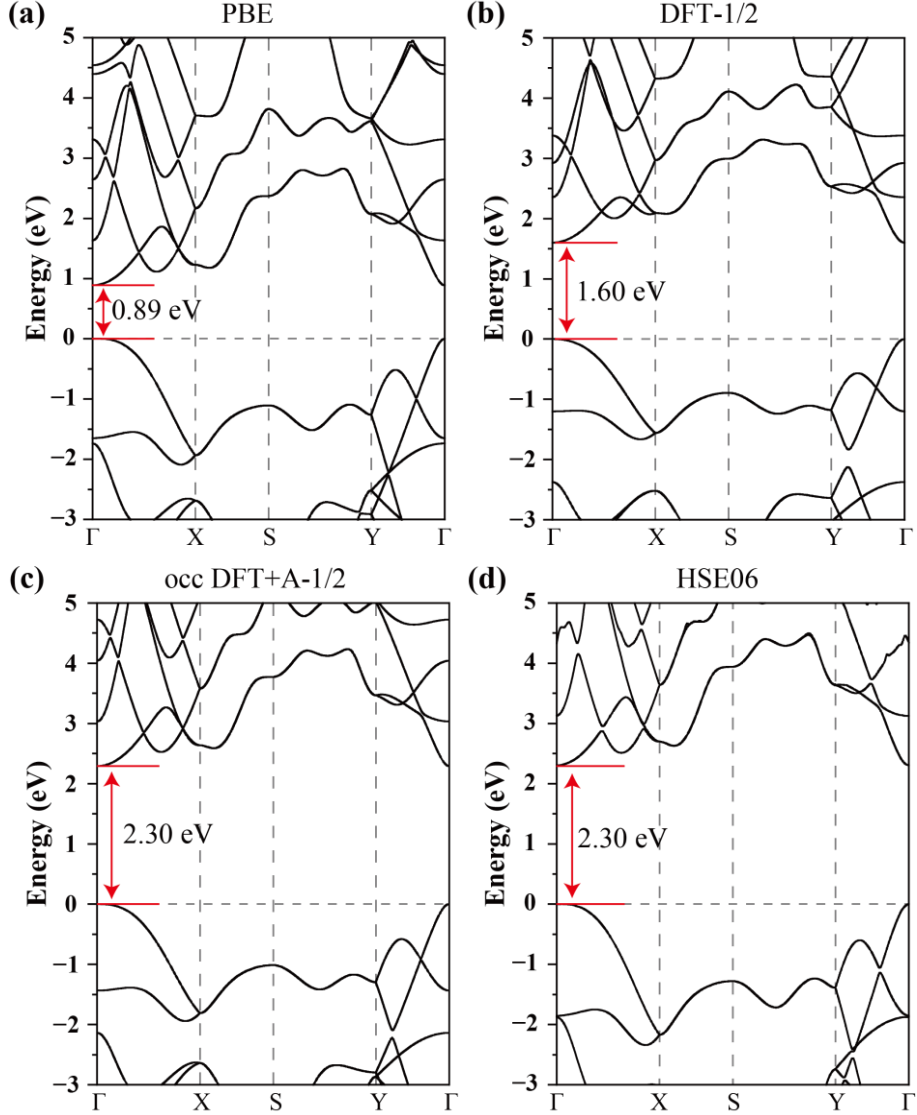

**Figure 6**. Band diagrams of monolayer BP, calculated through (a) PBE; (b) DFT-1/2; (c) occ DFT+*A*-1/2; (d) HSE06.

With the last argument in mind, we test the occ shell DFT+*A*-1/2 method, with comparison to HSE06 [31,32] hybrid functional calculations. Here the *A* factor acts on the SEPs as usual. **Figure 6** demonstrates the electronic band structures of monolayer BP, using GGA-PBE, DFT-1/2, occ DFT+*A*-1/2 and HSE06. DFT-1/2 already improves substantially compared with PBE, but the magnitude of the gap is still lower than the experimental value ~2 eV [33]. The HSE06 band gap is as large as 2.3 eV. Through tuning *A* to 0.33, occ DFT+*A*-1/2 also yields a 2.3 eV band gap. The conduction band structures are very similar between occ DFT+*A*-1/2 and HSE06, while the valence band is a bit compressed due to the -1/2 correction, which is a common phenomenon that has been pointed out before [13,15].

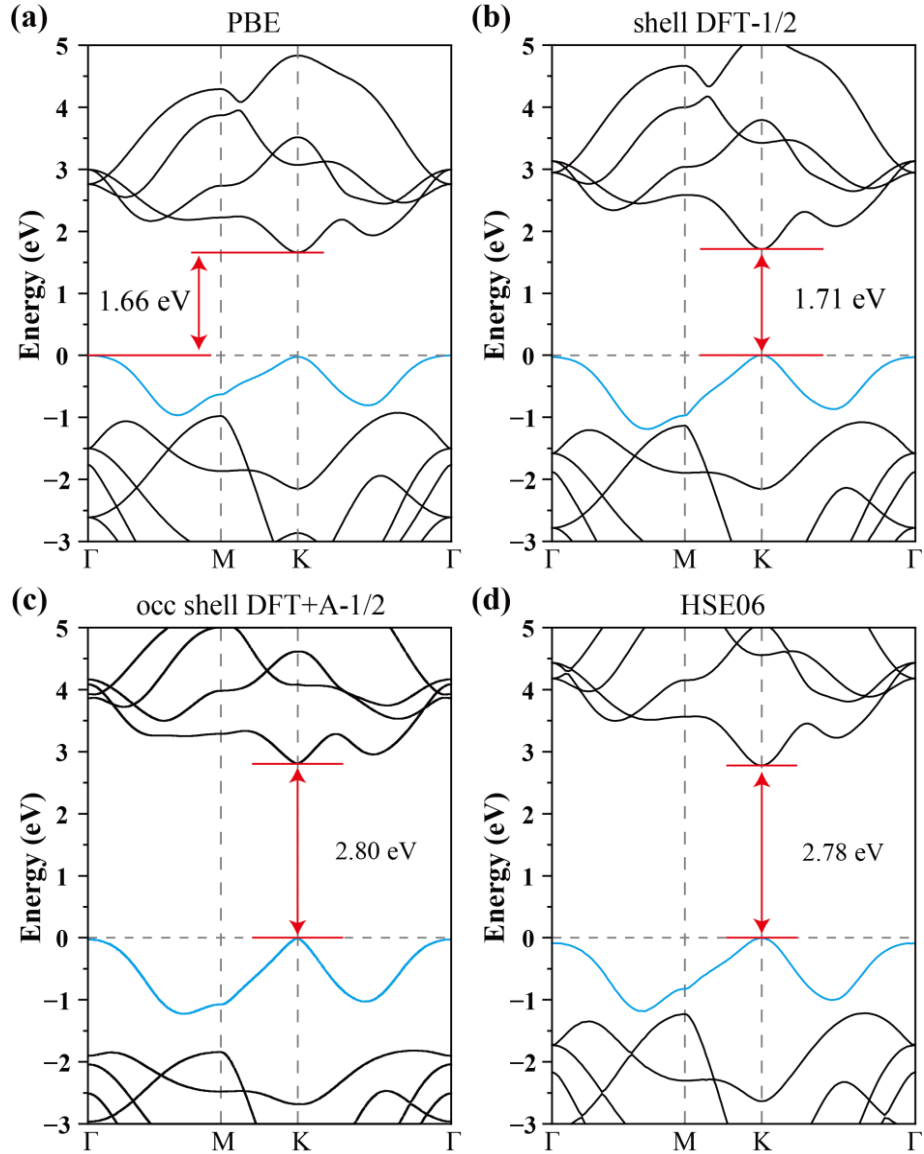

**Figure 7**. Band diagrams of monolayer MoS$_2$, calculated using (a) GGA-PBE; (b) DFT-1/2; (c) occ shell DFT+$A$-1/2; (d) HSE06.

In **Figure 7** we compared the calculated band diagrams of monolayer MoS$_2$. Shell DFT-1/2 correctly predicts the direct gap feature, though the improvement in the gap magnitude is very limited. In occ shell DFT+$A$-1/2, the A values are set to 0.84 for Mo and 0.32 for S. And it is predicted that monolayer MoS$_2$ is a $K-K$ direct gap semiconductor. There are moderate differences between the occ shell DFT+$A$-1/2 and HSE06 band structures. For example, the lowest conduction band eigenvalue at Γ is 3.87 eV above the VBM for occ shell DFT+$A$-1/2, but in HSE06 it is 4.18 eV. Moreover, the black valence bands are farther apart from the blue valence band in occ shell DFT+$A$-1/2. Yet, the overall band structure similarity between the two methods is observed.

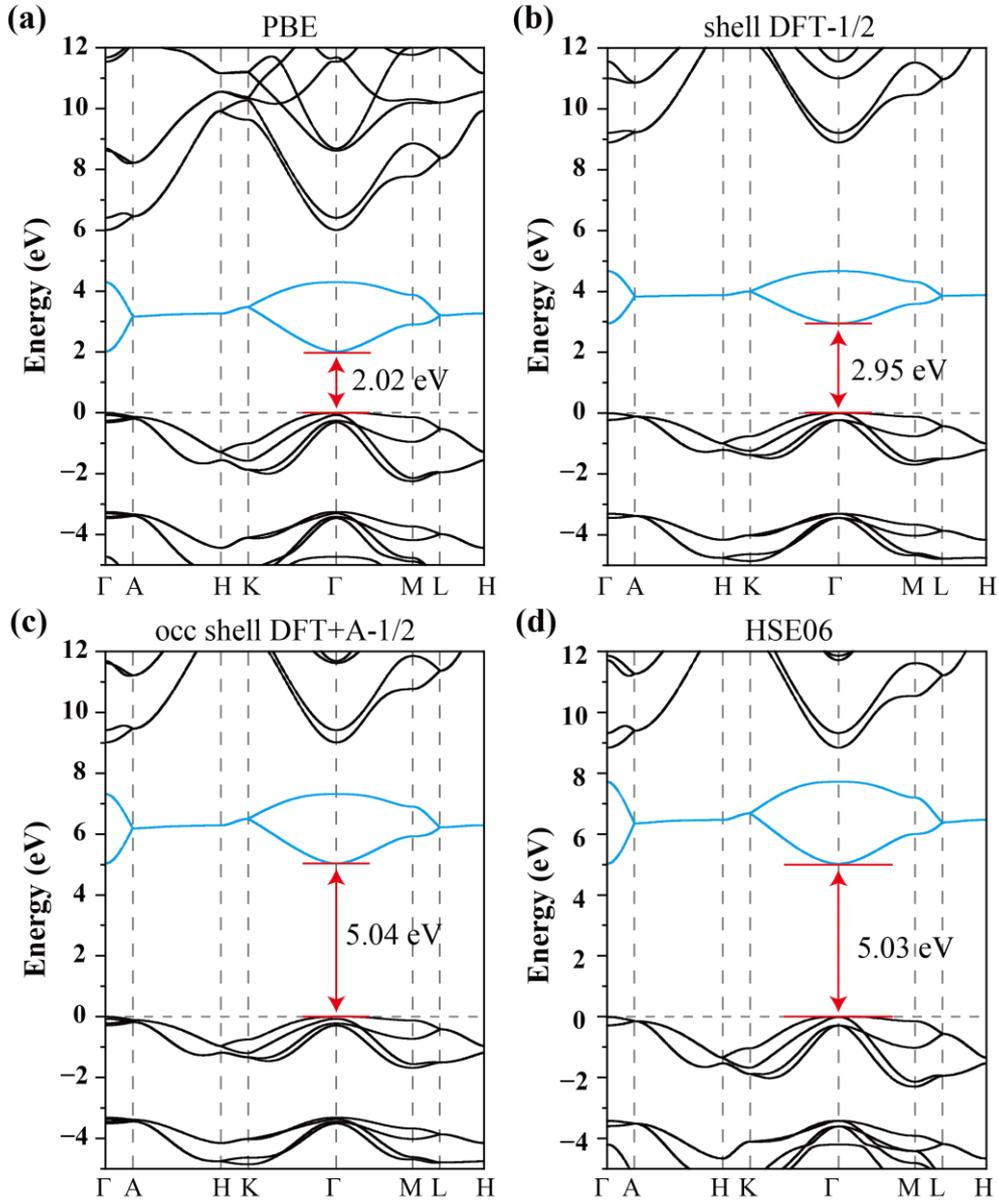

**Figure 8**. Band diagrams of $Li_2O_2$, calculated using (a) GGA-PBE; (b) DFT-1/2; (c) occ shell DFT+$A$-1/2; (d) HSE06.

$Li_2O_2$ is a typical example where shell DFT-1/2 fails, because O appears as $O^-$ such that both the CBM and the VBM are dominated by O states. As shown in **Figure 8**, shell DFT-1/2 only brings a tiny band gap increase compared with GGA. The blue conduction band is relevant to O states, which are pulled down too much in the -1/2 e correction, which ought to act on the valence band only. The even higher lying conduction bands (9 eV and above) consist of Li states, whose energies are nevertheless quite similar to HSE06 calculations. Hence, it is confirmed that the failure of shell DFT-1/2 stems from, as expected, the undesired downshift of the blue conduction bands. Therefore, occ shell DFT+$A$-1/2 is very suitable to overcome this limitation, and the resulting band diagram agrees well with that of HSE06. $Cu_2O$ is a similar case as $Li_2O_2$, since it involves $Cu^+$ with an intermediate +1 valency. Consequently, both the VBM and the CBM of $Cu_2O$ are predominantly composed of Cu states. This characteristic poses challenges for the application of the shell DFT-1/2 method. As demonstrated in **Figure 9**, the occ shell DFT+$A$-1/2 band diagram is satisfactory for $Cu_2O$, with $A = 0.5$ for Cu.

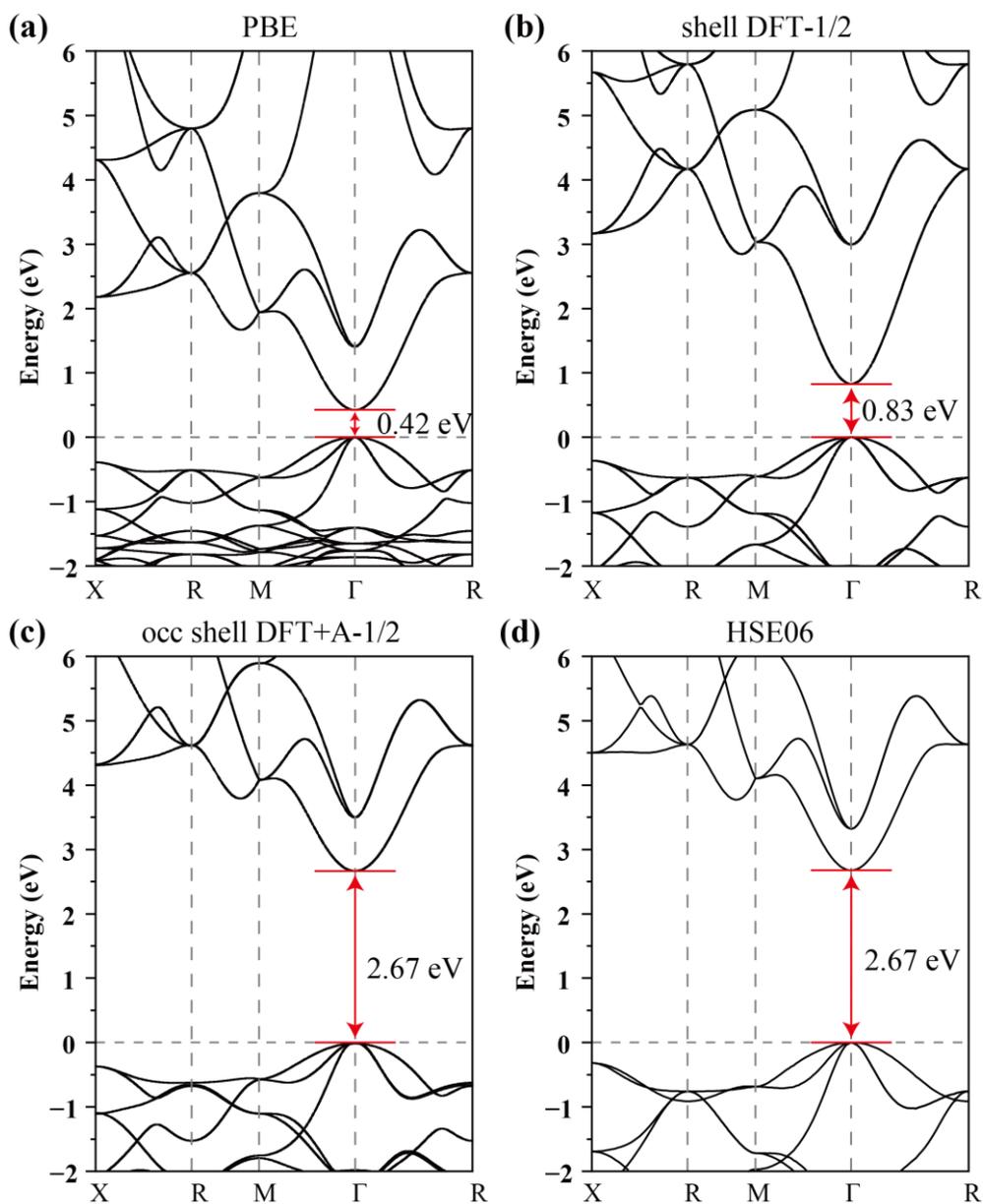

**Figure 9**. Band diagrams of $Cu_2O$, calculated using (a) GGA-PBE; (b) DFT-1/2; (c) occ shell DFT+$A$-1/2; (d) HSE06.

While occ DFT+$A$-1/2 could be tuned to yield reasonable band gap values, this semi-empirical method makes sense only if its absolute CBM and VBM levels are correct. A price is paid in this method, since it is no longer *ab initio*. Without additional gains, this price will be unnecessary. **Figure 10** illustrates the Hartree potential calculations with respect to vacuum level in 2D BP and $MoS_2$. The CBM and VBM values are readily available through comparison with the average Hartree potential. The detailed results are listed in **Table 1**. Compared with conventional GGA, both VBM and CBM are pulled down severely in BP, while the occ shell DFT+*A*-1/2 band edges are close to HSE06 results in terms of their absolute positions. For $MoS_2$, shell DFT-1/2 and GGA show similar results. Regarding the distance between the vacuum level and CBM, both PBE and HSE06 predict a value of ~4.3 eV, while occ shell DFT+*A*-1/2 predicts a smaller value. To make a fair judge, one has to refer to experimental values. The measured work function of monolayer $MoS_2$ is 5.77 eV [34], and the electron affinity is figured out to be 3.87 eV. With comparison to experimental, occ shell DFT+*A*-1/2 actually performs better than HSE06.

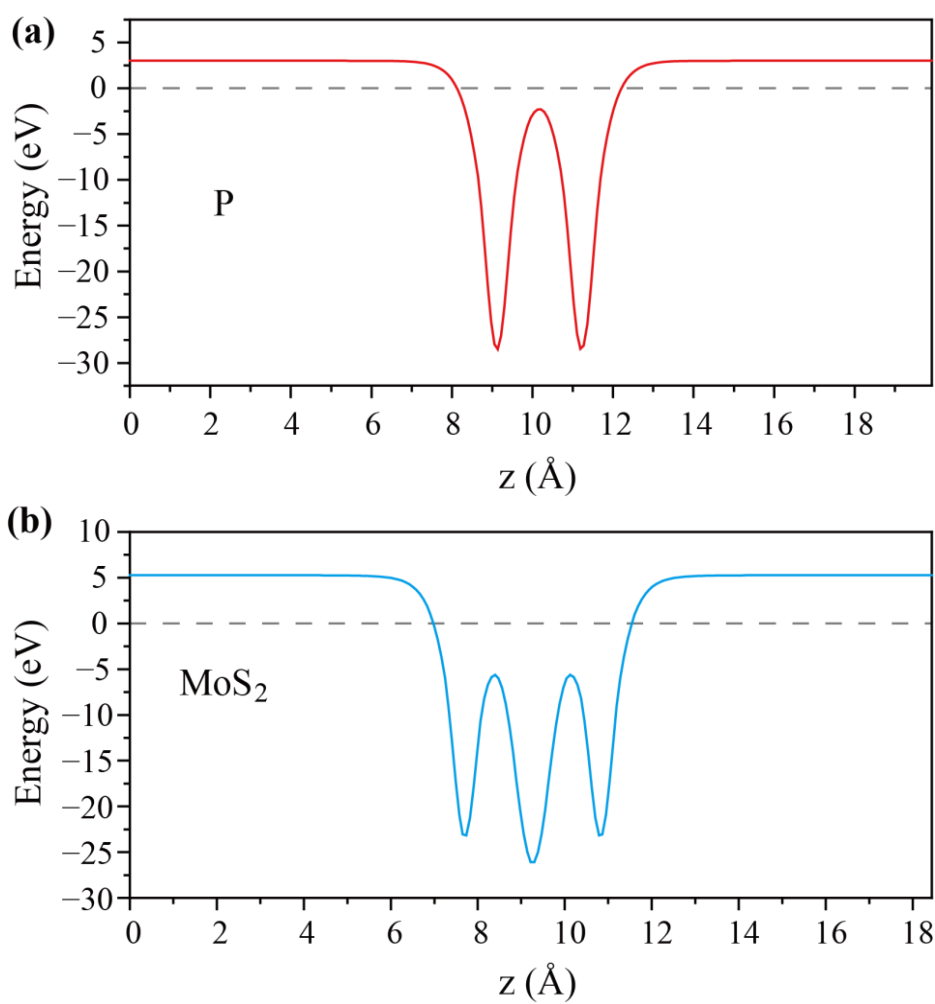

**Figure 10**. Average Hartree potentials calculated through occ shell DFT+*A*-1/2 in vacuum-containing 2D materials: (a) BP; (b) $MoS_2$.

**Table 1**. Distance between the vacuum level and VBM/CBM in monolayer BP and monolayer MoS$_2$ (unit: eV), calculated with various methods.

|                | PBE  |      | shell DFT-1/2 |      | Occ shell DFT+$A$-1/2 |      | HSE06 |      |
|----------------|------|------|---------------|------|-----------------------|------|-------|------|
|                | VBM  | CBM  | VBM           | CBM  | VBM                   | CBM  | VBM   | CBM  |
| P              | 5.04 | 4.15 | 7.54          | 5.94 | 6.38                  | 4.08 | 6.24  | 3.94 |
| MoS$_2$        | 5.94 | 4.28 | 5.97          | 4.26 | 6.79                  | 4.02 | 7.08  | 4.30 |

# V. CONCLUSION

In conclusion, we have tested the occupation dependent DFT-1/2 method (occ DFT-1/2) to overcome the undesired downshift in conduction band, which is common in standard DFT-1/2 and shell DFT-1/2. It forces the conduction band not to be disturbed by the self-energy potential that is designed for valence band correction. This method is particularly suitable to semiconductors with an intermediate valency element, or 2D materials. It is discovered that a shell-like self-energy potential region is still necessary to improve the exact band structure, and the inner and outer cutoff radii follow the shell DFT-1/2 values, but the occ shell DFT-1/2 band gaps are typically too large. This is because the self-energy potentials derived from atomic calculations are too strong for the delocalized electronic states in a solid. Through introducing the scaling factor $A$ for the self-energy potential, occ shell DFT+$A$-1/2 performs very well for monolayer black phosphorus, monolayer MoS$_2$, Li$_2$O$_2$ as well as Cu$_2$O. The occ shell DFT+$A$-1/2 electron affinity result for MoS$_2$ is superior to that of HSE06, though the computational load of HSE06 is at least two orders of magnitude higher than occ shell DFT+$A$-1/2.

## Acknowledgement

This work was supported by the National Natural Science Foundation of China under Grant No. 12474230.

# References

[1]   Sham L J and Schlüter M 1985 Density-functional theory of the band gap *Phys. Rev. B* **32** 3883–9

[2]   Hohenberg P and Kohn W 1964 Inhomogeneous Electron Gas *Phys. Rev.* **136** B864–71

[3]   Kohn W and Sham L J 1965 Self-Consistent Equations Including Exchange and Correlation Effects *Phys. Rev.* **140** A1133–8

[4]   Vosko S H, Wilk L and Nusair M 1980 Accurate spin-dependent electron liquid correlation energies for local spin density calculations: a critical analysis *Can. J. Phys.* **58** 1200–11

[5]   Perdew J P and Zunger A 1981 Self-interaction correction to density-functional approximations for many-electron systems *Phys. Rev. B* **23** 5048–79

[6]   Perdew J P and Wang Y 1992 Accurate and simple analytic representation of the electron-gas correlation energy *Phys. Rev. B* **45** 13244–9

[7]   Xie Q-X, Wu J and Zhao Y 2021 Accurate correlation energy functional for uniform electron gas from an interpolation ansatz without fitting parameters *Phys. Rev. B* **103** 045130

[8]   Perdew J P, Parr R G, Levy M and Balduz J L 1982 Density-Functional Theory for Fractional Particle Number: Derivative Discontinuities of the Energy *Phys. Rev. Lett.* **49** 1691–4

[9]   Levy M, Perdew J P and Sahni V 1984 Exact differential equation for the density and ionization energy of a many-particle system *Phys. Rev. A* **30** 2745–8

[10]  Almbladh C O and Pedroza A C 1984 Density-functional exchange-correlation potentials and orbital eigenvalues for light atoms *Phys. Rev. A* **29** 2322–30

[11]  Lonsdale D R and Goerigk L 2020 The one-electron self-interaction error in 74 density functional approximations: a case study on hydrogenic mono- and dinuclear systems *Phys. Chem. Chem. Phys.* **22** 15805–30

[12]  Baerends E J 2017 From the Kohn–Sham band gap to the fundamental gap in solids. An integer electron approach *Phys. Chem. Chem. Phys.* **19** 15639–56

[13]  Ferreira L G, Marques M and Teles L K 2008 Approximation to density functional theory for the calculation of band gaps of semiconductors *Phys. Rev. B* **78** 125116

[14]  Ferreira L G, Marques M and Teles L K 2011 Slater half-occupation technique revisited: the LDA-1/2 and GGA-1/2 approaches for atomic ionization energies and band gaps in semiconductors *AIP Advances* **1** 032119

[15]  Mao G-Q, Yan Z-Y, Xue K-H, Ai Z, Yang S, Cui H, Yuan J-H, Ren T-L and Miao X 2022 DFT-1/2 and shell DFT-1/2 methods: electronic structure calculation for semiconductors at LDA

complexity *J. Phys.: Condens. Matter* **34** 403001

[16] Cui H, Yang S, Yuan J-H, Li L-H, Ye F, Huang J, Xue K-H and Miao X 2022 Shell DFT-1/2 method towards engineering accuracy for semiconductors: GGA versus LDA *Computational Materials Science* **213** 111669

[17] Cui H, Yang S, Xue K-H, Huang J and Miao X 2023 On the self-consistency of DFT-1/2 *The Journal of Chemical Physics* **158** 094103

[18] Xue K-H, Yuan J-H, Fonseca L R C and Miao X-S 2018 Improved LDA-1/2 method for band structure calculations in covalent semiconductors *Computational Materials Science* **153** 493–505

[19] Yang S, Wang X, Liu Y, Wu J, Zhou W, Miao X, Huang L and Xue K-H 2022 Enabling *Ab Initio* Material Design of In As / Ga Sb Superlattices for Infrared Detection *Phys. Rev. Applied* **18** 024058

[20] Huang J, Yang W, Chen Z, Yang S, Xue K-H and Miao X 2024 Why Is the Bandgap of GaP Indirect While That of GaAs and GaN Are Direct? *physica status solidi (RRL) – Rapid Research Letters* **18** 2300489

[21] Doumont J, Tran F and Blaha P 2019 Limitations of the DFT–1/2 method for covalent semiconductors and transition-metal oxides *Phys. Rev. B* **99** 115101

[22] Anisimov V I, Zaanen J and Andersen O K 1991 Band theory and Mott insulators: Hubbard $U$ instead of Stoner $I$ *Phys. Rev. B* **44** 943–54

[23] Dudarev S L, Botton G A, Savrasov S Y, Humphreys C J and Sutton A P 1998 Electron-energy-loss spectra and the structural stability of nickel oxide: An LSDA+U study *Phys. Rev. B* **57** 1505–9

[24] Hubbard J 1963 Electron correlations in narrow energy bands *Proc. R. Soc. Lond. A* **276** 238–57

[25] Li P, Liu X, Chen M, Lin P, Ren X, Lin L, Yang C and He L 2016 Large-scale ab initio simulations based on systematically improvable atomic basis *Computational Materials Science* **112** 503–17

[26] Lin P, Ren X, Liu X and He L 2024 Ab initio electronic structure calculations based on numerical atomic orbitals: Basic fomalisms and recent progresses *WIREs Comput Mol Sci* **14** e1687

[27] Yang S, Xue K-H and Miao X 2025 Fundamentals of plane wave-based methods for energy band calculations in solids *J. Phys.: Condens. Matter* **37** 233001

[28] Wang H, Li C, Fang P, Zhang Z and Zhang J Z 2018 Synthesis, properties, and optoelectronic applications of two-dimensional $MoS_2$ and $MoS_2$-based heterostructures *Chem. Soc. Rev.* **47**

6101–27

[29] Ataide C A, Pelá R R, Marques M, Teles L K, Furthmüller J and Bechstedt F 2017 Fast and accurate approximate quasiparticle band structure calculations of ZnO, CdO, and MgO polymorphs *Phys. Rev. B* **95** 045126

[30] Ai Z, Yang S, Xue K-H, Yang W, Huang J and Miao X 2024 DFT-1/2 for ionic insulators: Impact of self-energy potential on band gap correction *Computational Materials Science* **239** 112978

[31] Heyd J, Scuseria G E and Ernzerhof M 2003 Hybrid functionals based on a screened Coulomb potential *The Journal of Chemical Physics* **118** 8207–15

[32] Heyd J, Scuseria G E and Ernzerhof M 2006 Erratum: "Hybrid functionals based on a screened Coulomb potential" [J. Chem. Phys. 118, 8207 (2003)] *The Journal of Chemical Physics* **124** 219906

[33] Margot F, Lisi S, Cucchi I, Cappelli E, Hunter A, Gutiérrez-Lezama I, Ma K, Von Rohr F, Berthod C, Petocchi F, Poncé S, Marzari N, Gibertini M, Tamai A, Morpurgo A F and Baumberger F 2023 Electronic Structure of Few-Layer Black Phosphorus from μ-ARPES *Nano Lett.* **23** 6433–9

[34] Keyshar K, Berg M, Zhang X, Vajtai R, Gupta G, Chan C K, Beechem T E, Ajayan P M, Mohite A D and Ohta T 2017 Experimental Determination of the Ionization Energies of $MoSe_2$, $WS_2$, and $MoS_2$ on $SiO_2$ Using Photoemission Electron Microscopy *ACS Nano* **11** 8223–30